\documentclass[aps,psfig,showpacs,groupedaddress]{revtex4}
\usepackage{color}
\usepackage{graphics}
\usepackage{epsfig}
\usepackage{ulem}

\newcommand{\bee}{\begin{equation}}
\newcommand{\ene}{\end{equation}}
\newcommand{\beea}{\begin{eqnarray}}
\newcommand{\enea}{\end{eqnarray}}

\begin{document}
\title{Magnetic field generation  in  finite beam plasma system}
\author{Amita Das$^1$}
\email{amita@ipr.res.in}
\author{Atul Kumar$^1$}
\author{Chandrasekhar Shukla$^1$}
\author{Ratan Kumar Bera$^1$}
\author{Deepa Verma$^1$} 
\author{Bhavesh Patel$^1$}
\author{Y. Hayashi$^2$}
\author{K. A. Tanaka$^3$}
\author{G. R. Kumar$^4$}
\author{Predhiman Kaw$^1$} 
\affiliation{$^1$Institute for Plasma Research, Bhat, Gandhinagar - 382428, India }
\affiliation{$^{2}$ Graduate School of Engineering, Osaka University, Suita, Osaka 565-0871 Japan }
\affiliation{$^{3}$  Extreme Laser Infrastructure-Nuclear Physics 30 Reactorului, PO Box MG-6, Bucharest Magurele 077125 Romania} 
\affiliation{$^{4}$ Tata Institute of Fundamental Research, 1 Homi Bhabha Road, Mumbai 400005, India} 

%\date{}
%{for revtex  maketitle should be written just here}
%\maketitle 
\begin{abstract} 
For   finite systems   boundaries can   introduce remarkable novel features. A well known example  is the 
Casimir effect \cite{PhysRev.73.360,PKNAW} that is observed in quantum electrodynamic systems. In classical systems too novel effects associated with 
finite boundaries have been observed,  for example  the surface plasmon mode  \cite{PhysRev.120.130} that appears when the  plasma has a finite extension. 
In this work a  novel instability 
associated with the  finite transverse size of a 
beam flowing through a plasma system has been  shown to exist.  This instability  leads to  
 distinct  characteristic  features of the  associated  magnetic field that gets  generated.   
For example, in contrast to the  well known  unstable Weibel mode of  a beam plasma system which generates  magnetic field at the skin depth scale,  
this  instability generates magnetic field at the scales length of 
 the transverse beam dimension \cite{chandra}. The existence of this new instability is demonstrated by  analytical arguments and by simulations conducted with the help of a variety of 
 Particle - In - Cell (PIC) 
codes (e.g. OSIRIS, EPOCH, PICPSI). Two fluid simulations have also been conducted 
which confirm the observations.  Furthermore, laboratory experiments  on laser plasma system  also  provides evidence 
of such an instability  mechanism at work. 

\end{abstract}
% for revtex4 here maketitle should be written
%\pacs{52.30.Cv,52.35.Ra} 
\maketitle 
%\begin{multicols}{2}
% upto here ------

\section{Introduction}
The dynamical evolution of magnetic field plays an important role in a variety of contexts ranging from astrophysical phenomena \cite{Remington} to laboratory plasmas. It is well known 
that when a high power laser impinges on an overdense  plasma target (and/or solid which can get ionized to form a plasma) it generates energetic electrons \cite{Modena, Brunela,PhysRevLett.81.822}. The 
current due to these energetic electrons are balanced by the return plasma current from the background electrons.  The combination of forward and return shielding currents is  
known  to be susceptible to the Weibel destabilization process as observed from 
the PIC simulations for periodic infinite beam - plasma system \cite{PhysRevE.65.046408}. 
Such a destabilization leads to spatially separated current filaments at the electron skin 
depth leading to magnetic field generation at commensurate scales. 
The choice of infinite periodic system 
is based on  an inherent understanding that the boundary effects due to 
finite system would merely have a small incremental impact.  This, however, turns out to be incorrect. 
 It has been shown here  with PIC as well as two fluid simulations that when a beam with finite 
 transverse extent is considered its   boundary introduces 
novel effects. In fact  an  entirely new instability associated with the finite size of the beam  
appears which generates  magnetic fields at the scale of the beam size right at the very outset. 
Secondly,  the  sheared electron flow configuration at the two edges of the finite beam 
is seen to be susceptible to Kelvin Helmholtz instability \cite{EMHD, cshukla}. Weibel \cite{Weibel} appears only in the bulk 
region of the beam. 
Thus, there are three  sources of magnetic fluctuations in a finite beam system - the new Finite 
Boundary System  (FBS) and KH instability
operating at the edge and the usual  Weibel destabilization process occurring in the  bulk 
region.  The theoretical description of the FBS instability  has been provided which is followed 
up by evidences from simulations (both PIC and fluid ) and experimental data\cite{Mondal}. 

We present a theoretical analysis of the new FBS instability which essentially confirms that the magnetic field appearance observed at the edge of the finite beam 
right in the beginning is due to a 
 new instability, which has so far not been identified. 
It is shown that the FBS instability, in fact, arises through the contribution from the boundary.  
An analytical understanding of the  characteristic features of the mode observed in simulations has  been provided. This particular mode has  direct impact on the long scale magnetic field 
formation at an early stage of beam plasma system. The experimental confirmation 
is provided  in terms of the spectral properties of the magnetic field produced by 
laser plasma interactions in the laboratory. The appearance of  magnetic fields at scales 
longer than the skin depth during very early stage 
 can only be  accounted for by this instability. The  
conventional Weibel destabilization route of the beam plasma system can account for such long scales much later in time when nonlinear 
inverse cascade effects have had an opportunity to produce 
 such effects.

The manuscript has been organized as follows. Section II contains the theoretical description of the new effect. 
In section III the simulation and experimental evidences  supporting the FBS effects have been provided. 
Section IV contains the conclusion. 

\section{Theoretical description}
An  equilibrium configuration  of the beam plasma system in 2-D $x-y$ plane is 
considered as shown in Fig.1.
The central region II from $-a \le y \le a$ carries the beam 
current and an oppositely flowing background plasma current which balance each other. In region 
I and region III the plasma is static 
and at rest. 
%We represent the beam and plasma electrons by suffices $b$ and $p$ respectively. 
The  charge  neutralization in equilibrium is achieved  
 by  balancing the total electron density by the 
 background ion density, viz., $\sum_{\alpha} n_{0\alpha} = n_{0i}$ in all the three regions.  
 The  electron flow velocity in region I and III is zero, whereas in region II it 
satisfies the condition of zero current, i.e. 
$ \sum_{\alpha} n_{0\alpha} v_{0\alpha x} = 0$. Here the suffix $\alpha $ stands for  
 $b$ for   beam 
electrons and  $p$ for the plasma electrons. 
The linearized perturbation of this equilibrium is considered with  variations in 
 $y$ and $t$ only. The flow is confined in $x- y$ plane, so  we have 
 $B_{1z}$, $E_{1x}$ and $E_{1y}$ (where the suffix $x,y,$ and $z$ denotes 
 the components) only as  the perturbed dominant fields.  
%The linearized continuity and the  momentum equations for the two fluids upon Fourier analyzing 
%in time yield: 
%\beea
%\frac{\partial v_{1\alpha y}}{\partial y} &=& i \omega \frac{n_{1\alpha}}{n_{0\alpha}} ; \\ 
%-i \omega \gamma_{0\alpha}^3 v_{1 \alpha x} + (\gamma_{0\alpha} v_{0\alpha x})^{\prime} 
%v_{1\alpha y} &=& -\frac{e E_{1x}}{m_0}; \\
%-i \omega \gamma_{0\alpha} v_{1\alpha y} &=& -\frac{e}{m_0} \left[ E_{1y} - \frac{1}{c} 
%v_{0x \alpha}B_{1z} \right]
%\label{lin-cont-mon}
%\enea
%The linearized Maxwell's equations give 
%\beea
%B_{1z} &=& \frac{ic}{\omega}\frac{\partial E_{1x}}{\partial y} \\
%\frac{\partial B_{1z}}{\partial y} &=& \frac{4 \pi}{c} J_{1x} - \frac{i \omega}{c} E_{1x} \\
%E_{1y} &=& \frac{4\pi e i}{\omega} \sum_{\alpha} n_{0 \alpha} v_{1\alpha y}
%\enea
%Here the superscript $\prime$ indicates differentiation with respect to $y$. 
%The  linearized perturbed current $J_{1x}$ has contributions from both  density perturbation and 
%the perturbed flow velocity. 
%Thus 
%\bee
%J_{1x} = -e \sum_{\alpha} n_{0 \alpha} v_{1 \alpha x} - e\sum_{\alpha} n_{1\alpha} v_{0 \alpha x}
%\ene
Eliminating all the perturbed fields in terms of $E_{1x}$ leads to the  following differential 
equation: 
\begin{equation}
\left[f_2 E_{1x}^{\prime} \right]^{\prime} - g_2 E_{1x} = 0
\label{finaleq}
\end{equation}
Here 
\begin{eqnarray}
f_2 &=& 1 + \frac{S_4}{\omega^2} - \frac{S_3^2}{\omega^2(S_1-\omega^2)} \\
g_2 &=& S_2 - \omega^2
\end{eqnarray}
where 
\begin{eqnarray}
S_1 &=& \sum_{\alpha} \frac{n_{0 \alpha}}{n_0 \gamma_{0 \alpha}}; \\
S_2 &=& \sum_{\alpha} \frac{n_{0 \alpha}}{n_0 \gamma_{0 \alpha}^3};\\
S_3 &=& \sum_{\alpha} \frac{n_{0 \alpha} v_{0 x \alpha}}{n_0 \gamma_{0 \alpha}}; \\
S_4 &=& \sum_{\alpha} \frac{n_{0 \alpha} v_{0 x \alpha}^2}{n_0 \gamma_{0 \alpha}};
\label{sdefs}
\end{eqnarray}
It should be noted that $S_3$ and $S_4$ are finite only when there is an equilibrium flow in the 
two fluid electron depiction. 
Furthermore, if the flow velocities of the two electron species are equal and opposite  then $S_3=0$.

The homogeneous limit of Califano $ \emph{et. al.} $\cite{Pegoraro,PhysRevE.58.7837} can be easily  recovered if we 
take Fourier transform of Eq.(\ref{finaleq}). The homogenous equation yields the dispersion relation 
for the  Weibel growth rate. 
We now seek the possibilities for obtaining purely growing modes in a finite system. 
For this purpose we multiply  Eq.(\ref{finaleq}) by $E_{1x}$, replace $\omega_2 = -\gamma^2$ (for 
purely growing modes) and integrate over $y$ over region II, i.e. from $-a$ to $a$. This yields: 
 \begin{equation}
\int_{-a}^a \left[ E_{1x} (f_2 E_{1x}^{\prime})^{\prime} - g_2 E_{1x}^2\right]dy = 0  \\
\label{inteq}
 \end{equation}
 Upon integrating by parts we obtain 
 \begin{equation}
f_2 \left[ E_{1x} E_{1x}^\prime \right]\vert_{-a}^{a} - \int_{-a}^{a} \left\lbrace f_2 \left[ E_{1x}^{\prime}\right]^2 +g_2 
E_{1x}^2 \right\rbrace dy = 0 \\
\label{inteq1}
 \end{equation}
In region II, $f_2$ and $g_2$ being constant, we can take them outside the integral. Thus Eq.(\ref{inteq1}) 
can be written as 
\begin{equation} 
f_2 \left[ E_{1x} E_{1x}^\prime \right]\vert_{-a}^{a} - f_2 \int_{-a}^{a}  \left[ 
E_{1x}^{\prime}\right]^2 dy  - g_2 
\int_{-a}^{a}E_{1x}^2  dy = 0 \\
\label{inteq2}
\end{equation}
If the boundary term is  absent, as in the case of infinite homogeneous system, the Eq.(\ref{inteq2}) 
can be 
satisfied for a finite value of $E_{1x}$ provided second and third terms have opposite signs. 
The integrand being positive definite this is possible provided $g_2$ and $f_2$ have opposite signs. 
The definitions of $g_2$ and $f_2 $ in terms of $\gamma^2$ are 
$$ f_2 = 1 + \frac{S_3^2}{\gamma^2(S_1+\gamma^2)} - \frac{S_4}{\gamma^2} $$
$$ g_2 = S_2 + \gamma^2$$
Since $g_2$ is  positive,  the only possible way for $f_2$ to be negative is to have 
$S_4/\gamma^2$ dominate over the first two terms of $f_2$. Thus, the conventional Weibel 
gets driven by $S_4$. It is also obvious that $S_3$ provides a stabilizing contribution making it 
more difficult for $f_2$ to become negative. A finite value of $S_3$ implies a non-symmetric flow 
configuration,  i.e. one for which the two  electron species have different flow speeds.  

For finite system something interesting happens when  boundary contributions are retained.  

The value of $\left[ E_{1x} E_{1x}^\prime \right]\vert_{-a}^{a}$ should be positive as 
 $E_{1x}^2$ should increase as one enters region II from region I (at $y = -a$) and it should decrease 
 at $y = a$. Thus the sign of first term will be  determined by the sign of $f_2$. 
 Another way to understand the positivity of  the sign of 
 $\left[ E_{1x} E_{1x}^\prime \right]\vert_{-a}^{a}$ 
 is by realizing that this term is essentially the radiative flux moving outside region II
 which can only be positive. This is seen  by casting it  in the form  of the Poynting flux by expressing the 
 derivative of $E_{1x}$ in terms of 
 $B_{1z}$. 

There exists the possibility then  that the first term 
of Eq.(\ref{inteq2}) has a finite contribution to balance the second and the third terms. 
Thus even if $f_2$ and $g_2$ have same signs (positive) Eq.(\ref{inteq2}) 
can be satisfied for a finite $E_{1x}$. It should be noted that the  
instability driven in this case is 
different from the Weibel mode as the boundary terms are responsible for it. 
It should be noted that Equation(\ref{inteq2}) can be 
satisfied most easily by the boundary contribution provided the variations in $E_{1x}$ in the bulk is 
minimal 
so as to have minimal (close to zero) contribution from the second term. Thus 
the instability driven by the boundary term would have a preference 
for long scale excitation. 

It should be noted   that 
contrary to the Weibel mode,   $S_3$   aids the FBS instability.  A finite and large 
$S_3$ may easily render $f_2$ to be positive which is required for this instability.

\section{Evidences}
\begin{figure}[h!]
\center
                \includegraphics[width=\textwidth]{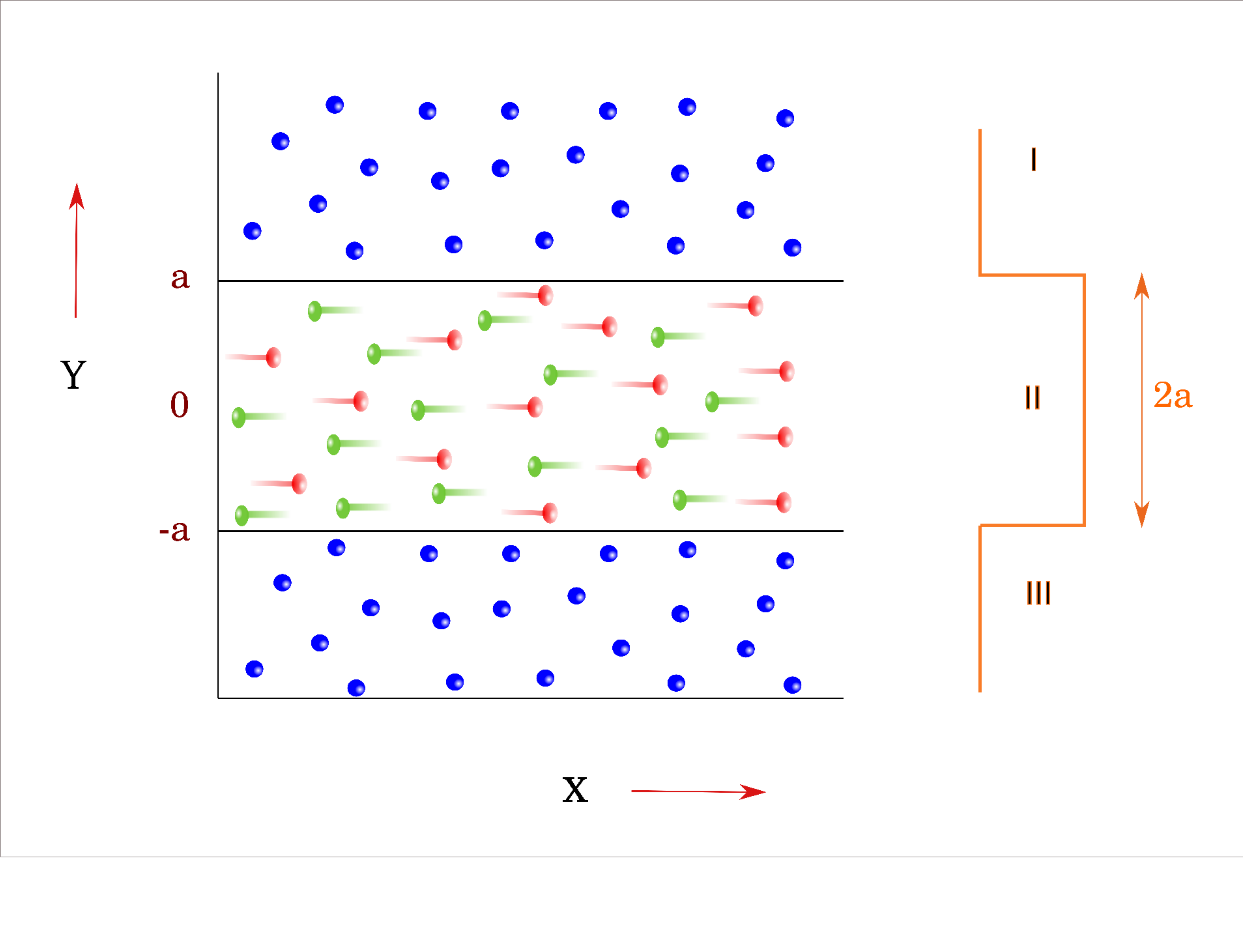} 
             \caption{ Schematics of 2D- equlibrium geometry of the beam plasma system where beam is finite in transverse direction   }  
                 \label{fig1}
         \end{figure}

\begin{figure}[h!]
  \center
                \includegraphics[width=\textwidth]{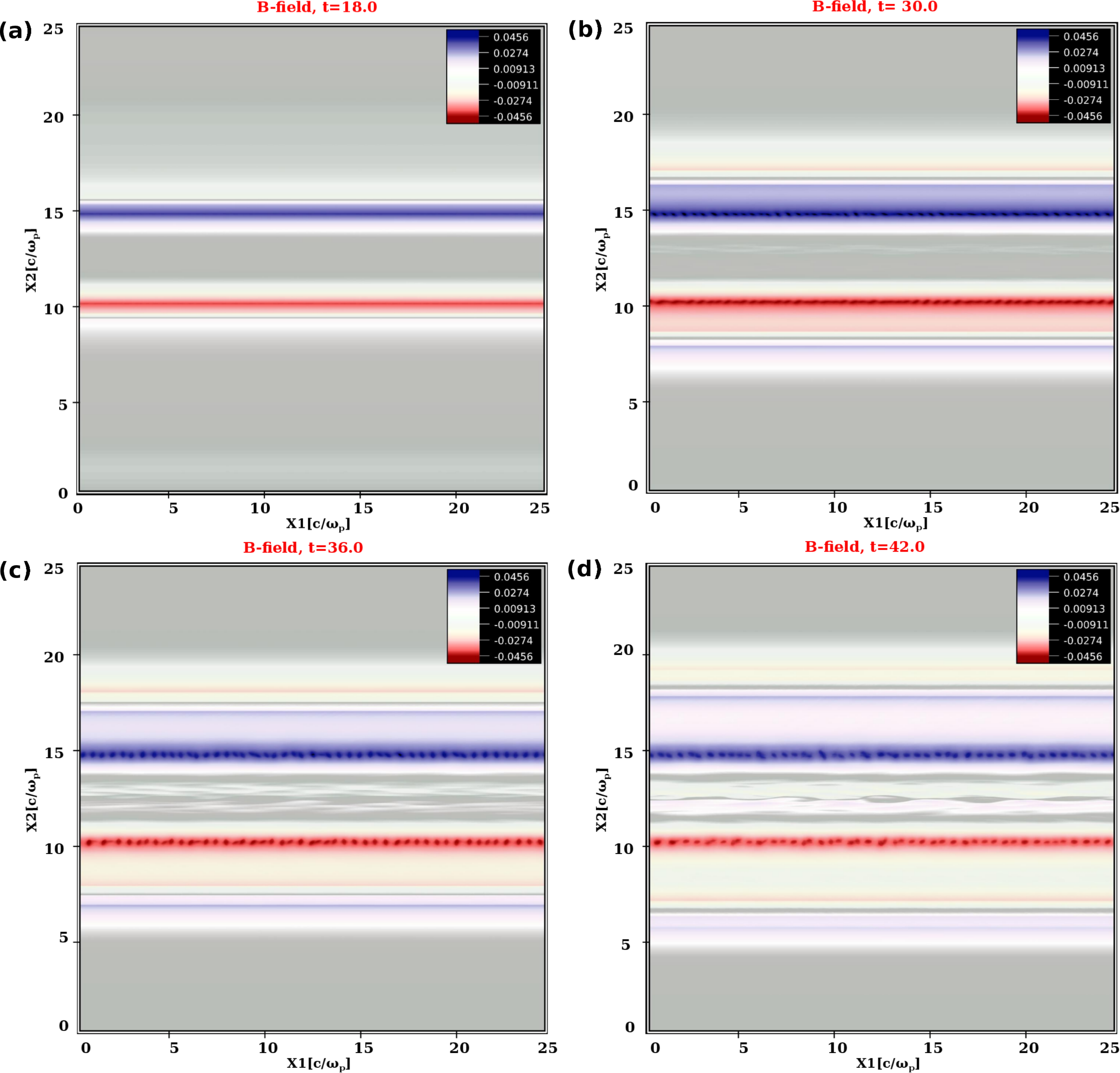} 
               \caption{ Snapshots of evolution of magnetic field B[in the units of $(m_e c\omega_p/e)$] with time t[in unit of $\omega_{pe}^{-1}$]:(a) At time $t=18.0$, Fig. shows the emergence of $B_z$ field with the opposite polarity
              at the edge of the beam (b) At time $t=30.0$, Fig. shows the initial devolopment of K-H vortices at the edges of the beam (c) Non-linear stage of K-H instability and the bulk region shows 
              the Weibel (current filamentation) instability.
              (d) Finally, the Magnetic field structures at the later time $t=42.0$ due to the presence of all three instabilities in the considered system. }  
                 \label{fig2}
         \end{figure}

\begin{figure}[h!]
  \center
                \includegraphics[width=\textwidth]{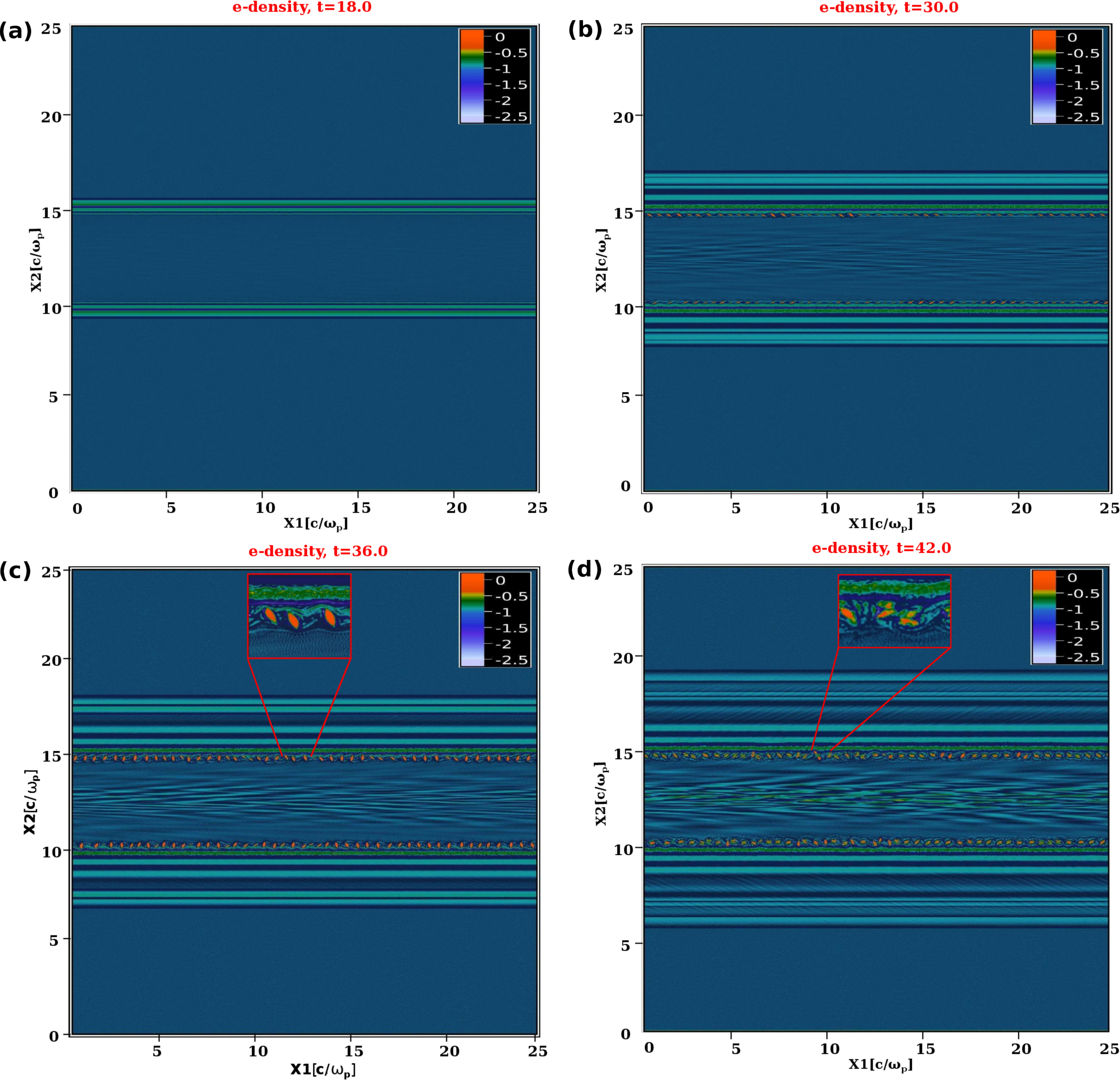} 
              \caption{ Snapshots of the evolution of electron density[in units of $n_0$] with time t[in the unit of $\omega_{pe}^{-1}$]:(a) The electron density configuration at initial time $t=18.0$ 
              (b) At time $t=30.0$, Fig. shows the initial devolopment of K-H vortices at the edges of the beam (c) the formation of K-H vortices at time $t=36.0$ at the edge of the beam which is highlighted by a red box and the bulk region shows 
             the Weibel (current filamentation) instability.
              (d) During the non-linear stage of K-H instabilities at time $t=42.0$, the fig. shows the merging of K-H vortices which is highlighted by a red box. }  
                 \label{fig3}
         \end{figure}

PIC simulations using OSIRIS \cite{Fonseca2002, osiris} and EPOCH \cite{epoch} were carried out for the case of a forward beam current and a compensating return plasma current 
of  a finite  extent,   at $t = 0$ shown as the  equilibrium configuration in Fig.~\ref{fig1}. 
The simulations were carried out and analyzed extensively   both 2-D and 3-D for various parameters. Here, for the sake of brevity we will present plots from our  2-D studies only. 
%In 3-D compensating cylindrical beam currents
% of diameter $2a$ was chosen. 
The simulation parameters for which the results are presented here correspond to  a 
   box size of 
$25 c/\omega_{pe} \times 25 c/\omega_{pe}$.  The beam is confined within 
an extent of $ 2a = 5 c/ \omega_{pe}$. Various choices of beam and background electron 
density have been studied. For instance, 
   beam electron density of $0.1 n_0$ moving  along $\hat{x}$ with a velocity of $0.9c$ 
   in a central region( from  $y = 10c/\omega_{pe}$ to $y = 15 c/\omega_{pe}$ ) 
i.e., with   a transverse extent of 
 $5 c/\omega_{pe}$ and  a  shielding current along $-\hat{x}$ of background electrons with density $0.9n_0$  moving with a 
  velocity of $0.1c$ was considered.   
  Here,  $n_0$ is the density of background ions which are at rest everywhere.  In the remaining region  from $y = 0$ to $10c/\omega_{pe}$ and $15$ to $20$ 
  $c/\omega_{pe}$
   electrons and ions both with density $n_0$ are at rest. Thus, the plasma everywhere is neutral with electron density balancing the density of 
  background plasma ions.  In the central beam region the beam current is exactly compensated by the return shielding current. 
 Snapshots at various 
  times for this particular case have been depicted  in Fig.~\ref{fig2} and Fig.~\ref{fig3} in the form of 2-D color plots for the $z$ component of the magnetic field and the  charge density respectively. 
%  \textcolor{red}{We also provide the snap shots from 3-D simulations for a finite beam propagating in the plasma Fig.~\ref{fig6} and Fig.~\ref{fig7}. 
%Here too all the three phases of evolution, viz., the FB followed by Kh and bulk Weibel mode 
%can be easily observed. }

\begin{figure}[h!]
  \center
                \includegraphics[width=\textwidth]{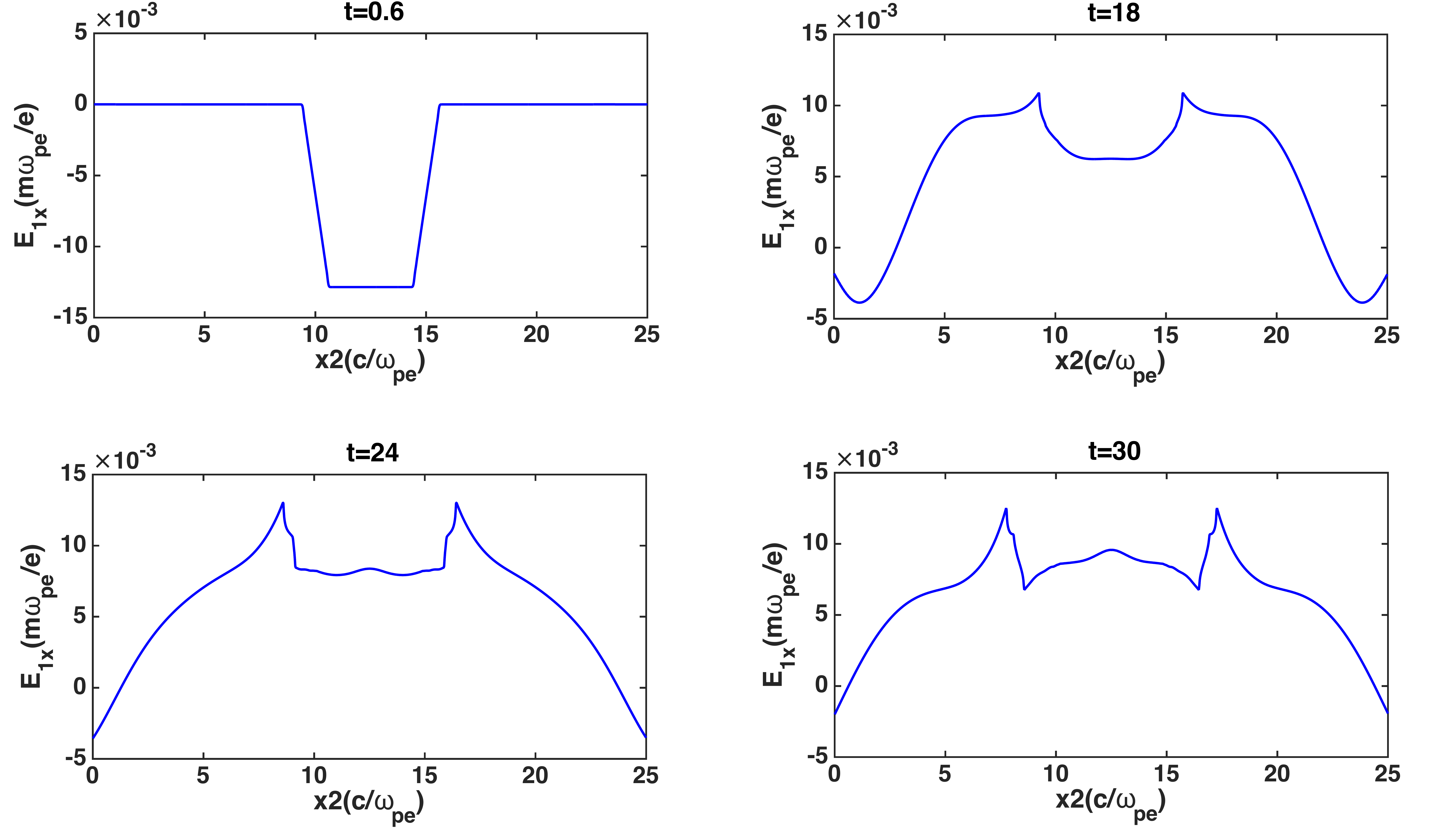} 
              \caption{Initial time evolution of $E_{1x}$  showing the minimal variation inside the beam region. }  
                 \label{fig4}
         \end{figure}

  From these plots it is clearly evident that there are  three distinct phases of evolution. 
  During the first phase 
 from $t = 0.12 {\omega_{pe}}^{-1}$ to $t = 30 {\omega_{pe}}^{-1} $ perturbed $z$ component of 
 magnetic field along $\hat{z}$ (transverse to 
 both flow and inhomogeneity)   appear  at the edges with 
  opposite polarity. This magnetic field has no $x$ dependence and is a function of $y$ alone. 
  This is consistent with the analytical choice.  
  The magnetic field perturbations are seen to grow with time and also 
  expand in $y$ from the edges in both the directions  at the speed of light. 
  The electron density perturbations, which also appear 
  at the edge,  on the other hand,  remain confined at the edge  during this phase. 
    This  first phase of the evolution, thus can   be characterized by the appearance 
     of  magnetic field perturbations with variations only along  $\hat{y}$, the transverse 
     direction. This suitably fits with the analytical description. Keeping in view that  
     the structures do not seem to vary with respect to the $x$,  the 1-D profiles along 
     $y$ have been shown in Fig.~\ref{fig4} for $E_{1x}$. 
     As predicted for the FBS mode theoretically,  $E_{1x}$ shows minimal variation inside the beam 
     region. The PIC simulations were also repeated for the case where $S_3 =0$ was taken 
     by a choice of symmetric flow. In this case we observed that the FBS mode did not appear. 
     This shows that $S_3$ plays a destabilizing role for this instability unlike is role 
     in the Weibel mode. 
   
During the second phase from $t = 30 {\omega_{pe}}^{-1} $  the Kelvin Helmholtz (KH) like 
perturbations 
appear at the edge of the current, and at a subsequently much later time viz., at around 
$t = 30 {\omega_{pe}}^{-1}$ one 
can observe the appearance of Weibel like perturbations in the bulk of the central 
region of the current flow. Both the KH and the Weibel mode 
have variations   along both $\hat{y}$ and $\hat{x}$ directions.  

%We also provide the snap shots from 3-D simulations for a finite beam propagating in the plasma. 
%Here too all the three phases of evolution, viz., the FB followed by KH and bulk Weibel mode 
%can be easily observed. 

These observations with characteristics three  phase 
developments have been  repeatedly observed in both 2-D and 3-D configurations  from a variety 
of  simulations carried out with different PIC  as well as fluid codes.  

\begin{figure}[h!]
  \center
                \includegraphics[width=\linewidth]{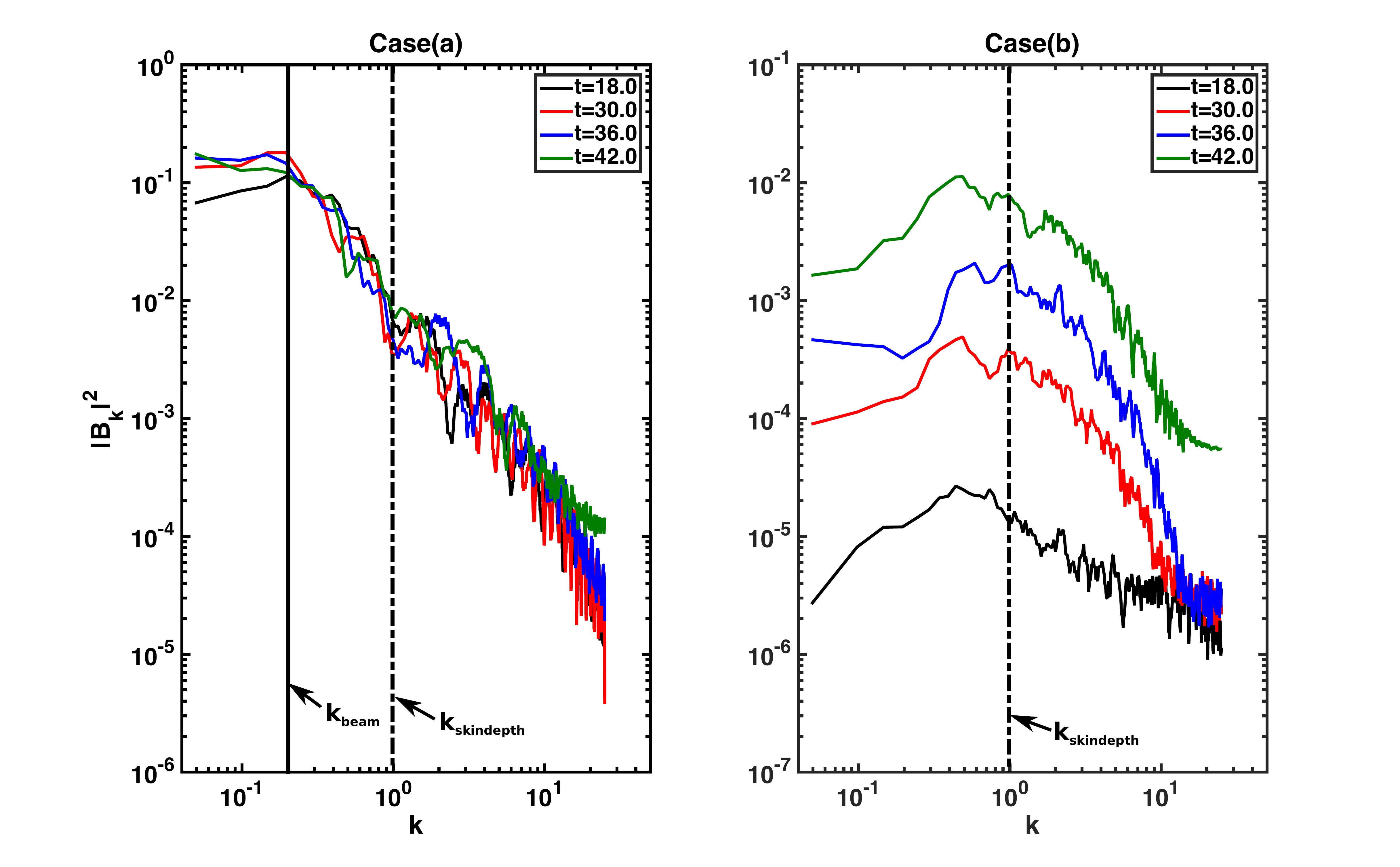} 
              \caption{Magnetic field spectra evolution with $ k $ : Case(a)-Finite beam-plasma system where field spectra peaks at the focal width of the beam; Case(b)-Infinite beam-plasma system where peak of the field spectra remains at the electron skin depth i.e. Weibel scale. }  
                 \label{fig5}
         \end{figure}

A comparison of the  
 evolution of the magnetic field spectra for the periodic infinite system as well as 
 the finite beam case has been shown in its snapshots 
 at various times in Fig.~\ref{fig5} . It is clear from the figure that for the periodic 
 infinite case the peak of  spectral power appears at the electron skin depth scale initially. 
 The spectral power only subsequently  cascades towards longer scales 
 via inverse  cascade mechanism.  On the other hand for the finite case it can be clearly observed 
 that the spectral peak appears at the beam size right from the very beginning. 
 In laboratory laser plasma experiments  the electron beam width would be 
 finite, typically commensurate with the dimension of the laser focal spot. The 
 measured magnetic 
 field  in a series of experiments where the electron beam gets generated at the critical 
 density surface and propagates in the overdense plasma medium, clearly show  that right from 
 the very beginning viz. t=0.12 $ \omega_{pe}^{-1} $ time scales the spectra maximizes at the longest measurable scale 
 and not the electron  skin depth scale  as one would expect from a Weibel like destabilization 
 process. These experiments \cite{Mondal} provide yet another evidence for the existence of a FBS instability.

\section{Summary}
While the KH and the Weibel modes are well known and have been discussed extensively in 
the literature, the FBS mode has neither been observed and nor been described anywhere. 
We thus report the first observation of this mode which relies entirely on the systems with  finite 
boundary effects.  The implications of this particular instability on magnetic field generation 
needs to be evaluated in different contexts. For instance, it is likely that the 
finite size  jets emanating from astrophysical objects are susceptible to  
this particular instability.  This work also suggests that the finite size considerations 
in many other systems need to be looked afresh to unravel new effects which might 
have been overlooked so far. \\

\bibliographystyle{ieeetr}  

\bibliography{finite_arxiv}

\begin{thebibliography}{10}

\bibitem{PhysRev.73.360}
H.~B.~G. Casimir and D.~Polder, ``The influence of retardation on the
  london-van der waals forces,'' {\em Phys. Rev.}, vol.~73, pp.~360--372, Feb
  1948.

\bibitem{PKNAW}
H.~B.~G. Casimir, ``On the attraction between two perfectly conducting
  plates,'' {\em Proc. K. Ned. Akad. Wet.}, vol.~51, pp.~793--795, 1948.

\bibitem{PhysRev.120.130}
E.~A. Stern and R.~A. Ferrell, ``Surface plasma oscillations of a degenerate
  electron gas,'' {\em Phys. Rev.}, vol.~120, pp.~130--136, Oct 1960.

\bibitem{chandra}
C.~Shukla, A.~Das, and K.~Patel, ``Effect of finite beam width on current
  separation in beam plasma system: Particle-in-cell simulations,'' {\em
  preprint arXiv.org}, 2015.

\bibitem{Remington}
A.~B. Remington, A.~David, P.~R., Drake, and T.~Hideaki, ``Modeling
  astrophysical phenomena in the laboratory with intense lasers,'' {\em
  Science}, vol.~284, no.~5419, pp.~1488--1493, 1999.

\bibitem{Modena}
A.~Modena, Z.~Najmudin, A.~Dangor, C.~Clayton, K.~A. Marsh, C.~Joshi, V.~Malka,
  C.~Darrow, C.~Danson, D.~Neely, and F.~N. Walsh, ``Electron acceleration from
  the breaking of relativistic plasma waves,'' {\em Nature}, vol.~377,
  pp.~606--608, Sep 1995.

\bibitem{Brunela}
F.~Brunel, ``{Anomalous absorption of high intensity subpicosecond laser
  pulses},'' {\em Physics of Fluids}, vol.~31, no.~9, p.~2714, 1988.

\bibitem{PhysRevLett.81.822}
K.~B. Wharton, S.~P. Hatchett, S.~C. Wilks, M.~H. Key, J.~D. Moody,
  V.~Yanovsky, A.~A. Offenberger, B.~A. Hammel, M.~D. Perry, and C.~Joshi,
  ``Experimental measurements of hot electrons generated by ultraintense
  laser-plasma interactions on solid-density targets,'' {\em Phys. Rev. Lett.},
  vol.~81, pp.~822--825, Jul 1998.

\bibitem{PhysRevE.65.046408}
Y.~Sentoku, K.~Mima, Z.~M. Sheng, P.~Kaw, K.~Nishihara, and K.~Nishikawa,
  ``Three-dimensional particle-in-cell simulations of energetic electron
  generation and transport with relativistic laser pulses in overdense
  plasmas,'' {\em Phys. Rev. E}, vol.~65, p.~046408, Mar 2002.

\bibitem{EMHD}
A.~Das and P.~Kaw, ``Nonlocal sausage-like instability of current channels in
  electron magnetohydrodynamics,'' {\em Physics of Plasmas}, vol.~8, no.~10,
  pp.~4518--4523, 2001.

\bibitem{cshukla}
C.~Shukla, A.~Das, and K.~Patel, ``Particle-in-cell simulation of
  two-dimensional electron velocity shear driven instability in relativistic
  domain,'' {\em Physics of Plasmas}, vol.~23, no.~8, p.~082108, 2016.

\bibitem{Weibel}
E.~S. Weibel, ``Spontaneously growing transverse waves in a plasma due to an
  anisotropic velocity distribution,'' {\em Phys. Rev. Lett.}, vol.~2,
  pp.~83--84, Feb 1959.

\bibitem{Mondal}
S.~Mondal, V.~Narayanan, W.~J. Ding, A.~D. Lad, B.~Hao, S.~Ahmad, W.~M. Wang,
  Z.~M. Sheng, S.~Sengupta, P.~Kaw, A.~Das, and G.~R. Kumar, ``Direct
  observation of turbulent magnetic fields in hot, dense laser produced
  plasmas,'' {\em Proceedings of the National Academy of Sciences}, vol.~109,
  no.~21, pp.~8011--8015, 2012.

\bibitem{Pegoraro}
F.~Pegoraro, S.~V. Bulanov, F.~Califano, and M.~Lontano, ``{Nonlinear
  development of the weibel instability and magnetic field generation in
  collisionless plasmas},'' {\em Physica Scripta}, vol.~T63, no.~3,
  pp.~262--265, 1996.

\bibitem{PhysRevE.58.7837}
F.~Califano, R.~Prandi, F.~Pegoraro, and S.~V. Bulanov, ``Nonlinear
  filamentation instability driven by an inhomogeneous current in a
  collisionless plasma,'' {\em Phys. Rev. E}, vol.~58, pp.~7837--7845, Dec
  1998.

\bibitem{Fonseca2002}
R.~A. Fonseca, L.~O. Silva, F.~S. Tsung, V.~K. Decyk, W.~Lu, C.~Ren, W.~B.
  Mori, S.~Deng, S.~Lee, T.~Katsouleas, and J.~C. Adam, {\em OSIRIS: A
  Three-Dimensional, Fully Relativistic Particle in Cell Code for Modeling
  Plasma Based Accelerators}, pp.~342--351.
\newblock Berlin, Heidelberg: Springer Berlin Heidelberg, 2002.

\bibitem{osiris}
R.~A. Fonseca, S.~F. Martins, L.~O. Silva, J.~W. Tonge, F.~S. Tsung, and W.~B.
  Mori, ``One-to-one direct modeling of experiments and astrophysical
  scenarios: pushing the envelope on kinetic plasma simulations,'' {\em Plasma
  Physics and Controlled Fusion}, vol.~50, no.~12, p.~124034, 2008.

\bibitem{epoch}
T.~D. Arber, K.~Bennett, C.~S. Brady, A.~Lawrence-Douglas, M.~G. Ramsay, N.~J.
  Sircombe, P.~Gillies, R.~G. Evans, H.~Schmitz, A.~R. Bell, and C.~P. Ridgers,
  ``Contemporary particle-in-cell approach to laser-plasma modelling,'' {\em
  Plasma Physics and Controlled Fusion}, vol.~57, no.~11, p.~113001, 2015.

\end{thebibliography}

\end{document}